\documentclass[10pt,emptycopyrightspace]{ewsn-proc}

\usepackage[]{xcolor}

\usepackage{graphicx}
\usepackage{balance}
\usepackage{comment}
\usepackage{url}

\numberofauthors{4}

\author{
\alignauthor Artis Rušiņš \\
    \affaddr{Institute of Electronics and Computer Science}\\
    \affaddr{Dzerbenes str. 14, Riga, Latvia}\\
    \email{artis.rusins@edi.lv}
\alignauthor Krišjānis Nesenbergs \\
    \affaddr{Institute of Electronics and Computer Science}\\
    \affaddr{Dzerbenes str. 14, Riga, Latvia}\\
    \email{krisjanis.nesenbergs@edi.lv}
\and
\alignauthor Deniss Tiščenko \\
    \affaddr{Institute of Electronics and Computer Science}\\
    \affaddr{Dzerbenes str. 14, Riga, Latvia}\\
    \email{deniss.tiscenko@edi.lv}
\alignauthor Pēteris Paikens \\
    \affaddr{Institute of Mathematics and Computer Science}\\
    \affaddr{University of Latvia}\\
    \affaddr{Raina blvd. 29, Riga, Latvia} \\
    \email{peteris.paikens@lumii.lv}
}

\title{An experimental study: RF Fingerprinting of Bluetooth devices}

\begin{document}

\maketitle

\begin{abstract}
This paper presents an experimental study on radio frequency (RF) fingerprinting of Bluetooth Classic devices. Our research aims to provide a practical evaluation of the possibilities for RF fingerprinting of everyday Bluetooth connected devices that may cause privacy risks. We have built an experimental setup for recording Bluetooth connection in a radio frequency isolated environment using commercially available SDR (software defined radio) systems, extracted fingerprints of the Bluetooth radio data in the form of carrier frequency offset and scaling factor from 6 different devices, and performed k-nearest neighbors (kNN) classification achieving 84\% accuracy. The experiment demonstrates that no matter what privacy measures are being taken in the protocol layer, the physical layer leaks significant information about the device to unauthorized listeners. In the context of the ever-growing Bluetooth device market, this research serves as a clarion call for device manufacturers, regulators, and end-users to acknowledge the privacy risks posed by RF fingerprinting and lays a foundation for more sizeable Bluetooth fingerprinting analysis research.
\end{abstract}

%
%

%

\section{Introduction}
  \label{sec:intro}
Bluetooth devices are ubiquitous in our society, as the annual Bluetooth device shipments worldwide stood at 4.9 billion units in 2022 and yearly shipments are forecast to reach 7.6 billion units in 2027 \cite{statista_wearables}, the impact of any privacy or security risks they might pose is potentially very high.

With the advances in technology, devices are becoming more sophisticated and can collect a wide range of data about the user - location, sensor data, connection metadata, biometric data, personalized settings, and others. The exchange of data typically occurs between various devices, such as smartphones, wearable devices, laptops, and peripheral devices, but the thing that does not change is that they usually use popular Bluetooth or Bluetooth Low Energy (BLE) standards which we are investigating. The devices are also ubiquitous enough that the activities of many people can be tracked by detecting the presence of a particular Bluetooth device.

The purpose of this research is to evaluate the feasibility of identifying unique Bluetooth devices without relying on information that is broadcast on higher Bluetooth abstraction layers like broadcasted device name or its MAC address. RF fingerprinting is one of the emerging techniques that can identify a specific device or type/model of the device by analyzing radio waveforms generated by the device under test and extracting unique features from it. Because these unique features arise from physical imperfections of Bluetooth radio chip they are impossible to fix by using any protocol layer anonymization features like MAC address randomization.

Fingerprinting can make attacks on devices more personalized thus it is important to investigate its effectiveness before offering countermeasures. However, Bluetooth device RF fingerprinting is a relatively unexplored research topic, and the practical aspects of extracting unique fingerprints using commercially available off-the-shelf devices like Software Defined Radios (SDR) remain a significant area of investigation. Additionally, we aim to determine the precision and accuracy of these extracted fingerprints in order to assess the viability of RF fingerprinting as a means of device identification.

To accomplish this, the research methodology involves a review of RF fingerprinting literature and a practical implementation for extracting multiple fingerprint types from intercepted radio packets between the Device Under Test (DUT) and another communication party. Interception is done using SDRs in a radio frequency-isolated environment. The precision and accuracy of the fingerprinting are evaluated after device identification using the kNN classification algorithm.

\section{Background}
In order to provide context on the analyzed process and experimental setup, it is first necessary to discuss the core concepts of Bluetooth communications and the privacy-related parts of the standard. We also introduce the concept of RF fingerprinting and describe various fingerprinting techniques found in the literature.

\subsection{Bluetooth Classic and Bluetooth Low Energy}
Bluetooth is a continuously evolving standard catering to different use cases, branching into two main options: Bluetooth Classic and Bluetooth Low Energy. Bluetooth operates within the unlicensed industrial, scientific, and medical (ISM) frequency band, spanning from 2.402 to 2.480 GHz.

Bluetooth Classic, also known as Bluetooth Basic Rate/Enhanced Data Rate (BR/EDR), is a low-power communication standard primarily used for audio streaming and data transfer. On the physical layer, it utilizes 79 channels with a bandwidth of 1 MHz each. In contrast, Bluetooth Low Energy (BLE) was introduced as part of the Bluetooth 4.0 standard and is specifically designed for low-energy use cases such as Internet of Things (IoT), smart home devices, and wearables. It operates within the same frequency range but utilizes 40 channels, each with a bandwidth of 2 MHz. Both variations of Bluetooth employ a master-slave architecture, where one master device coordinates communication with slave devices. Bluetooth uses Frequency-Hopping Spread Spectrum (FHSS) which essentially means that communication happens on one channel at a time, but the chosen channel is rapidly changing. The order of channels used or "hopping sequence" is derived from the internal clock of the master device so it is a very low probability that some other Bluetooth connection would interfere. Also, this sequence is unknown to third parties \cite{BluetoothCoreSpecv5.4} \cite{bluetooth_specification}. 
\par
Each Bluetooth device has a Bluetooth Device Name (the user-friendly name) - a UTF-8 string, which it exposes to remote devices without authentication or authorization. When the user enables Bluetooth scanning on a smartphone, this name is listed. Additionally, every Bluetooth device has a Bluetooth Address (BD\_ADDR), which is similar to a MAC address used for internet-connected devices. The BD\_ADDR is a 48-bit value that serves as a unique identifier for the Bluetooth device. It can be either a public device address or a random device address, depending on the choice made by the device developers. The device name can usually also be changed by end users.
\par 
The public device address ranges need to be registered with the Institute of Electrical and Electronics Engineers (IEEE), but a more prevalent type of address used is the random device address. When establishing a connection with a device utilizing a random device address, an Identity Resolving Key (IRK) is exchanged with a trusted device. This IRK assists in resolving the true BD\_ADDR from the broadcasted random device address. The purpose of this mechanism is to mitigate tracking risks associated with static addresses and to enhance user privacy \cite{BluetoothCoreSpecv5.4}. For example, on an iPhone, the devices listed under "Bluetooth, My Devices" consider this phone as a trusted device.
\par

\begin{figure*}[!ht]
    \centering
    \includegraphics[width=0.8\paperwidth]{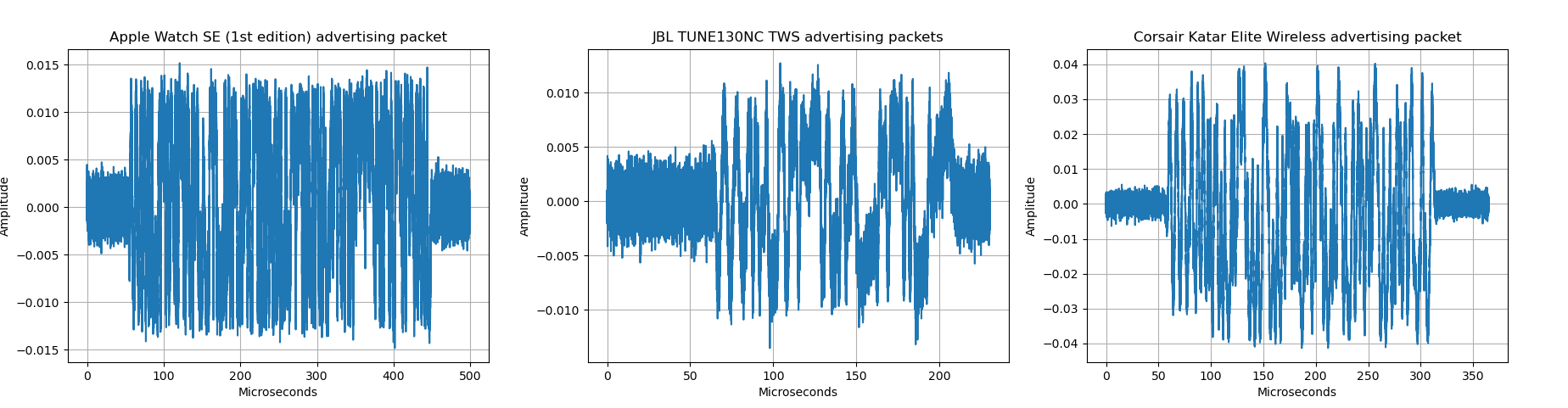}
    \caption{Advertising data packets for 3 Bluetooth devices in radio-isolated environment}
    \label{fig:3adv_packets}
\end{figure*}

\subsection{RF fingerprinting}
Fingerprinting in the context of electronic devices and their cyber security is a method where devices are distinguished and identified using a compilation of their unique features and characteristics. In this research, we deal with RF fingerprints. We look at the physical waveforms generated by radio chipsets without interpreting them as protocol fields for the Bluetooth protocol stack since parts of the standard that would allow us to identify the transmitting device - address, and name, can be changed by either developer and/or user and thus are not reliable identifications.
\par

One method to identify a device is by analyzing its generated signal ramp-up or ramp-down time, i.e., transient signal, which happens at every generated waveform while the transmitter turns on or off. These transient signals exhibit distinct characteristics that can be used to differentiate between different devices. Various methods exist to determine the start and stop times of the transient signal for individual packets: Bayesian Step Change Detection, Variance Fractal Dimension Threshold Detection, and others \cite{soltanieh2020review}. Existing research has shown that this method can yield very high classification accuracy, however, it also requires a recording device with a very high sampling rate such as a digital oscilloscope. The use of a high sampling rate is essential to capture the fine details and rapid changes in the transient signals accurately. In a study conducted by Kose et al. \cite{kose2019rf} researchers used a sample rate of 5 GSamples/s which would generate very large amounts of data. Since we plan to publish our dataset at a later time, sharing the entire large dataset can be unreasonable from a logistical standpoint and also prohibits potential third parties from adding their own data to this dataset, as prohibitively expensive equipment would be required in order to do so.

While carrier frequencies that should be used by Bluetooth devices are defined in the specification, in practice, the actual generated signal may have some unwanted frequency offset due to imperfections in the internal clock of the transmitter. In the context of RF fingerprinting the carrier frequency offset (CFO) can be used as an additional characteristic. Even so, it is important to note that the frequency offset is highly dependent on oscillator temperature, and embedded devices like smartphones are particularly exposed to temperature variation because of their dense internal components and limited cooling capabilities. The study conducted by Givehchian et. al \cite{givehchian2022evaluating} has shown that it is possible to extract the CFO of BLE packets collected with a low-cost SDR and in combination with In-Phase/Quadrature (IQ) offset this can generate accurate RF fingerprints. Although the IQ offset is a fingerprint that is only present in devices that use combined chipsets for multiple communication standards like Bluetooth and Wi-Fi, Givehchian et. al has shown that CFO which is present in every Bluetooth chipset, is a viable component of the fingerprint.
\par
There are studies that have explored the topic of Bluetooth device indoor localization by analyzing the Received Signal Strength Indication (RSSI) of received packets and utilizing a fingerprinting technique to estimate the location of the transmitter \cite{pei2010using}. We aim to complement this knowledge that RSSI can identify devices and our practical findings on how different signal amplitudes appear for various devices in the same environment as seen in Figure \ref{fig:3adv_packets}. In practical scenarios, variations in signal amplitude, along with Carrier Frequency Offset (CFO), need to be corrected to ensure the signal falls within the desired amplitude range of approximately [-1;1], before passing it to the demodulator to ensure a higher probability of receiving correct bits. The way to normalize signals amplitude to the required amplitude range is by multiplying it with some constant. This \textit{scaling factor} constant which takes into account all variations in amplitude can also be used as an RF fingerprint.

\section{Threat model}
There are several privacy and security concerns that arise from the possibility of fingerprint extraction from Bluetooth-enabled devices. The most obvious threat is related to unwanted tracking of the device itself or the person or vehicle carrying it without the consent of the device owner. More devious attack vectors include the identification of a specific Bluetooth device model or even hardware version in order to exploit device-specific vulnerabilities. A related threat to the identification of specific device models carried by unsuspecting persons involves more targeted attacks that involve the knowledge about the device model or capabilities, such as spear phishing using customized prompts that refer to the specific accessory device being used or even preparing physical in-person attacks on the device, that require knowledge of the device model used, such as plugging in infected memory sticks or replacing it with an identically looking altered model (e.g. compromised Bluetooth mouse or keyboard).

On the other hand, identification of the carried Bluetooth devices and related known vulnerabilities could allow the development of more effective protection measures for controlling entry at secure facilities with restrictions to specific types or capabilities of wireless devices.

\section{Experimental setup}
Our first goal is to record a dataset of Bluetooth device communication while mitigating interference from other devices in the crowded ISM band. To achieve this, we do all recordings in an anechoic radio frequency isolated chamber with SDRs and then perform RF fingerprint extraction, and classification later. We record all phases of communication -- advertising, communication establishment, data streaming, and disconnect. Our available test devices are listed in Table \ref{tab:device_list}.

\begin{table}[h!t]
\caption{List of test devices}
\renewcommand*\arraystretch{1.2}
\begin{center}
\begin{tabular}{|c|c|}
\hline
\textbf{Device} & \textbf{Type} \\
\hline
Apple Watch SE (1st gen.) & Smartwatch\\
\hline
Corsair Katar Elite Wireless & Wireless mouse\\
\hline
RF Wireless controller & Wireless presenter remote\\
\hline
JBL Tune 130NC TWS & Earbuds\\
\hline
Soundcore Liberty Air 2 Pro & Earbuds\\
\hline
Redmi Buds 3 & Earbuds \\
\hline
\end{tabular}
\label{tab:device_list}
\end{center}
\end{table}

    \subsection{Hardware}
    This fingerprinting approach requires the capability to record all possible 79 (or 40) Bluetooth channels at the same time, which requires recording 80 MHz of bandwidth and is a challenge for most  popular low-cost SDR models. While there are SDRs capable of capturing this bandwidth (for example, a popular company Ettus Research offers a solution at around 10k EUR \cite{ettus}), for cost reasons we decided to capture the traffic with a combination of 2 Ettus Research B210's, where each one of them will capture 40 MHz of spectrum, resulting in a more affordable solution.
    \par
    Given that many Bluetooth devices require a physical button press to start the connection and transmission process, conducting this action while the device is inside a closed shielded chamber presents challenges. To overcome this, we designed and made a "robot hand". This mechanical device securely holds the Bluetooth device and employs servo motors controlled by an Arduino board through an optically isolated USB connection to press an appropriate physical button on the device.
    \par
    To establish a connection with DUT we still require another communication party and for this, we use the Samsung Galaxy S20 FE smartphone which is controlled with Android Debugging Bridge (ADB) \cite{adb}. This bridge allows us to enable Bluetooth on the smartphone, start the pairing process with DUT, stream data for some time, and then disconnect without physically interacting with the smartphone. Instead, commands are passed through optically isolated USB, providing greater consistency of experiments through the use of control scripts.
    \par
\begin{figure}[h!t]
    \centering
    \includegraphics[width=0.4\paperwidth]{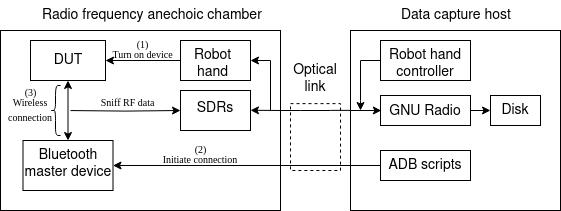}
    \caption{Data capture setup with 3 steps of the experiment. During (1), (2), and (3) everything is recorded by SDRs}
    \label{fig:data_capture_block}
\end{figure}
\begin{figure}[h!t]
    \centering
    \includegraphics[width=0.35\paperwidth]{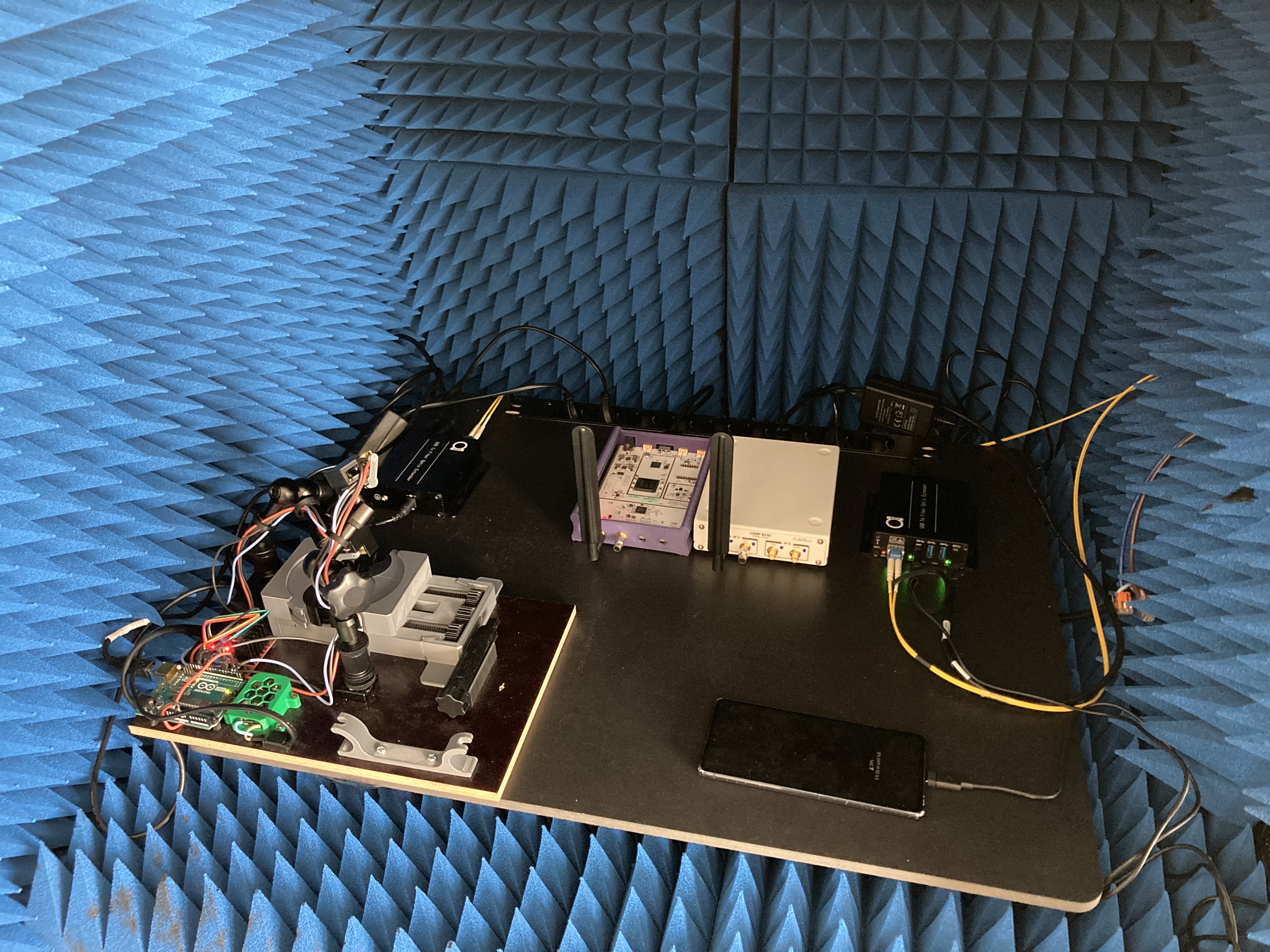}
    \caption{Data capture setup}
    \label{fig:data_capture_hw}
\end{figure}

    All the USB connections including two SDRs, the Android smartphone, and "robot hand" are connected to a USB-to-optical converter which is then connected to another optical-to-USB converter outside the chamber and then connected to a data collection PC. Optical wires minimize the risk of picking up or transmitting unwanted radio signals which could happen if we used regular copper USB cables. Figures \ref{fig:data_capture_block} and \ref{fig:data_capture_hw} show the hardware setup.

    \subsection{Software}
    To record radio data we use GNU Radio software with Ettus Research USRP blocks \cite{usrp_source} with configuration parameters described in Table \ref{tab:sdr_params}. This way we cover all possible Bluetooth channels. Through this hardware and software setup, we can capture Bluetooth device communication in a controlled environment, allowing for subsequent analysis.

\begin{table}[h!t]
\renewcommand*\arraystretch{1.2}
\centering
\caption{SDR parametrs}
\begin{tabular}{|c|c|c|} \hline
 & \textbf{USRP B210 (1)} & \textbf{USRP B210 (2)}\\ \hline
Center frequency & 2421.5 MHz & 2461.5 MHz\\ \hline
Sample rate & 40 $\cdot 10^6$ & 40 $\cdot 10^6$\\ \hline
Gain & 30 dB & 30 dB\\ \hline
Output data format & complex64 & complex64\\ \hline
Output file & radio\_lower.data & radio\_upper.data\\
\hline\end{tabular}
\label{tab:sdr_params}
\end{table}

    \par
\begin{figure}[h!t]
    \centering
    \includegraphics[width=0.4\paperwidth]{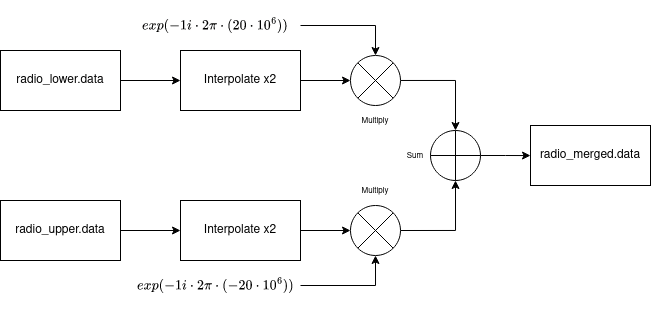}
    \caption{Merging of both data streams}
    \label{fig:radio_merged}
\end{figure}
\begin{figure}[h!t]
    \centering
    \includegraphics[width=0.4\paperwidth]{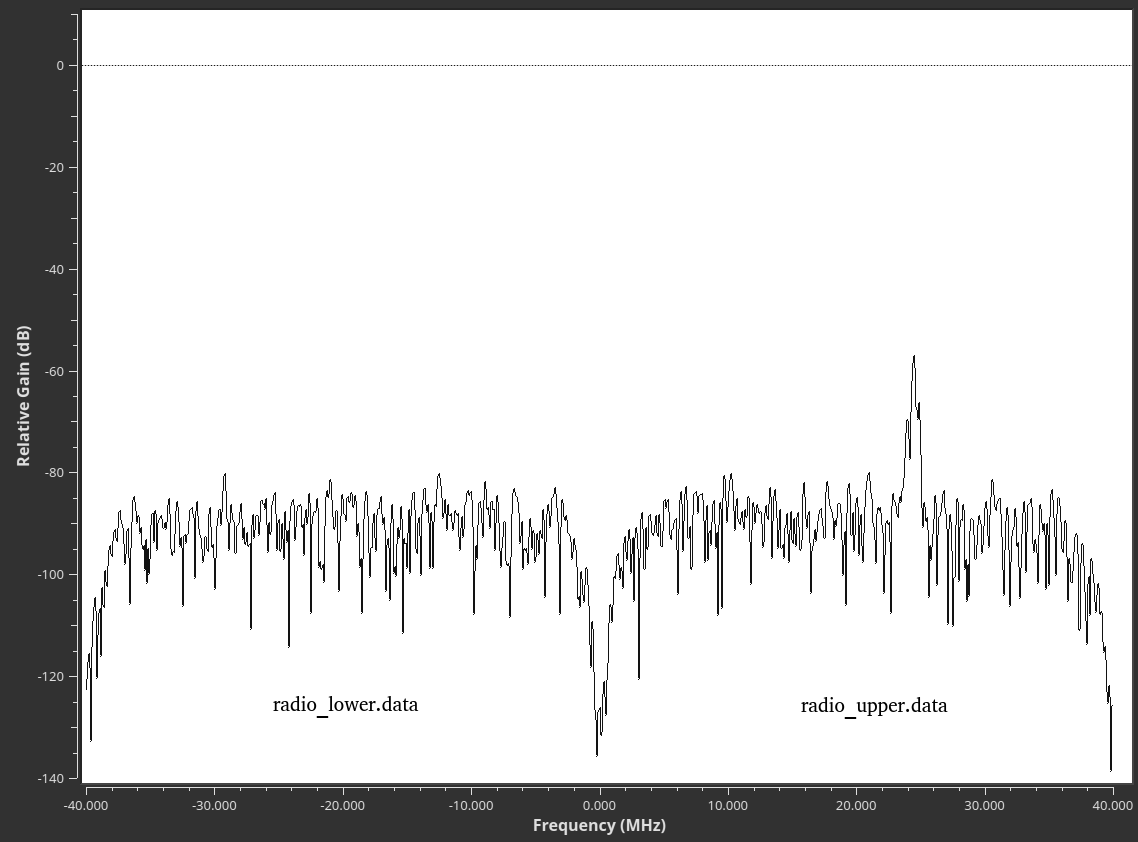}
    \caption{Frequency spectrum of radio\_merged.data}
    \label{fig:radio_merged_fft}
\end{figure}

    Because Bluetooth uses FHSS, every data packet can fall into either radio recordings and for further analysis, we combine both data streams into one. The first data file captured the frequency band between 2401.5 MHz and 2441.5 MHz and the second one the band between 2441.5 MHz and 2481.5 MHz. First, we interpolate both data streams with a factor of 2 to match the required sample rate of 80 MHz, next, each interpolated data stream was shifted by multiplying it with a sine wave of 20 MHz and -20 MHz frequency, respectively and finally, both streams are summed together resulting in radio\_merged.data file which captures frequencies from 2401.5 to 2481.5 MHz with a sample rate of 80 MHz. The process of data stream merging can be seen in Figure \ref{fig:radio_merged}. The frequency spectrum of the merged data file can be seen in Figure \ref{fig:radio_merged_fft}. 
    There are dips in the middle of the spectrum because of filtering done by SDR for signals outside the original 40 MHz bandwidth 
    but nevertheless, this does not limit our ability to analyze individual packets 
    because no Bluetooth channel falls exactly in the middle of the spectrum but instead aligns with the borders of the filter.

    \par
    Next, it is needed to detect and extract individual packets out of the merged data stream by performing energy detection, tagging detected packets and FHSS dehopping using Sandia Labs GNU Radio out-of-tree module \cite{gr-fhss}. Dehopping is realized by multiplying detected packets with their closest Bluetooth channel frequency. The output of this process is an individual file for each data packet at the baseband with some noise before and after it. Although slower than the often-used channelizing method we opted for the energy detection method because channelized data streams are decimated which could potentially negatively affect the quality of RF fingerprints. Then each packet is low pass filtered to remove noise and from filtered packets CFO and scaling factor is extracted. The algorithm utilized for this purpose is derived from the work of Mike Ryan from ICE9 Consulting LLC \cite{ice9}. They used this calculation to normalize the packet before passing it to the demodulator, but we save these values as RF fingerprints. For each detected, filtered Bluetooth packet:
    \begin{enumerate}
    \item Separate all samples of negative and positive amplitude
    \item $max = median(positive\_samples)$
    \item $min = median(negative\_samples)$
    \item $CFO = (max - min) / 2$
    \item $scaling\_factor = (max - CFO) / 2$
    \end{enumerate}
We repeat all of these processing steps for all of the test devices and save the extracted fingerprint data - CFO and scaling factor.

\section{Results}
\begin{figure}[h!t]
    \centering
    \includegraphics[width=0.35\paperwidth]{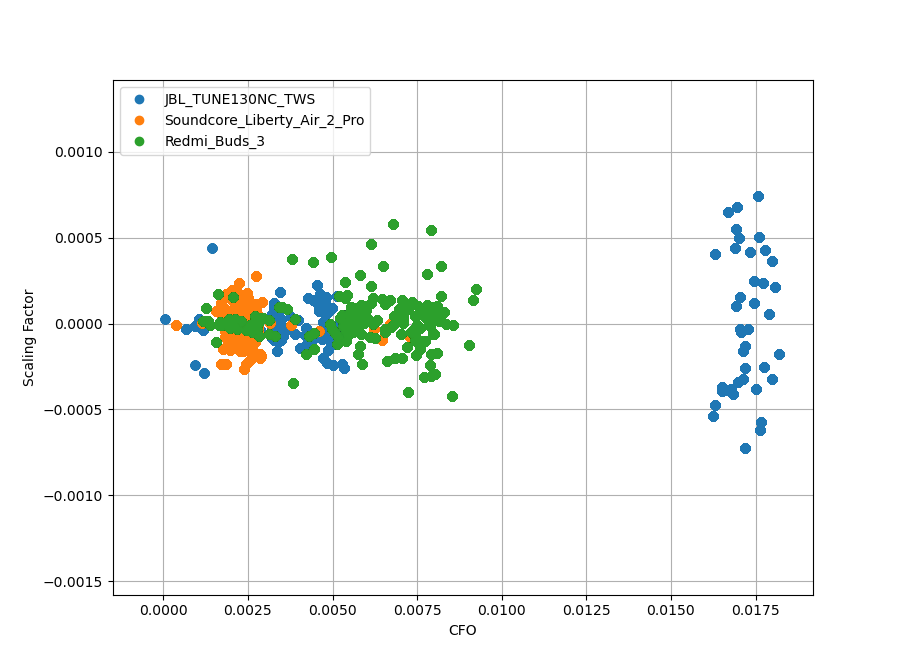}
    \caption{CFO vs Scaling factor scatter plot}
    \label{fig:cfo_sf_initial}
    \end{figure}

    Although we have a data set of the whole communication procedure, for initial results we only use the part where DUT is advertising and don't include parts where the smartphone connects to it because then we would need to filter out excess packets sent by the smartphone. Figure \ref{fig:cfo_sf_initial} shows CFO vs Scaling Factor scatter plot for 3 devices and it can be visually seen that the data points are distinguishable between devices therefore our RF fingerprinting method is possible at least for the selected test devices.

\begin{figure*}[!ht]
    \centering
    \includegraphics[width=0.7\paperwidth]{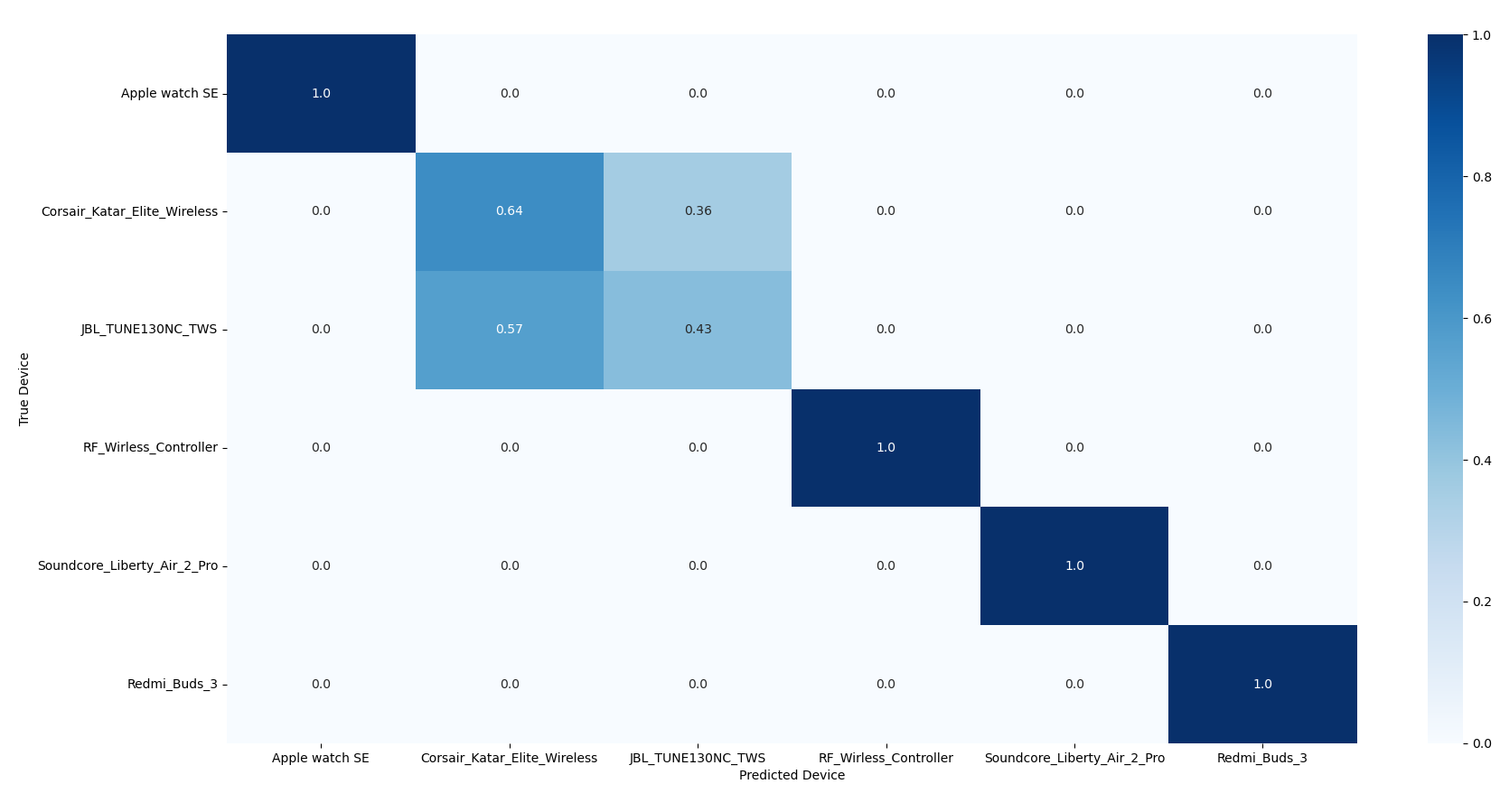}
    \caption{Normalized confusion matrix for all test devices}
    \label{fig:knn}
\end{figure*}

\par
For quantitative evaluatation the precision of captured RF fingerprint accuracy we apply kNN classification for all extracted fingerprints of 6 test devices. For this task, kNN was chosen as there is a limited number of clearly distinguishable clusters, and for fingerprinting use cases that enables attempting a classification even from a single captured packet, matching it to known device samples. The number of neighbors was chosen as k = 10, and 20\% of data was used as test data and the rest of it as training data. Figure \ref{fig:knn} shows the normalized confusion matrix for all extracted fingerprints of 6 test devices. Table \ref{tab:knn} shows evaluation metrics of the kNN classificator. Classifier achieved 84\% accuracy, precision, recall, and F1 score. It is important to note that the only 2 devices that could be confused are Corsair Katar Elite wireless mouse and JBL Tune 130NC TWS earbuds, every other device is classified with 100\% accuracy.

\begin{table}[h!t]
\centering
\renewcommand*\arraystretch{1.2}
\caption{k-NN evaluation}
\begin{tabular}{|c|c|} \hline
Accuracy & 0.8424 \\ \hline
Precision & 0.8429  \\ \hline
Recall & 0.8424 \\ \hline
F1 score & 0.8407 \\
\hline\end{tabular}
\label{tab:knn}
\end{table}


\section{Discussion and conclusions}
In this study, we recorded a dataset of 6 Bluetooth test devices in an isolated and controlled environment, capturing the whole communication process with commercially available Ettus Research SDRs, without any significant RF interference. We then extracted individual Bluetooth packets from the entire dataset and further extracted CFO and scaling factor from each individual packet as an RF fingerprint. We verified that extracted fingerprints could be used to track these individual devices with accuracy/precision/recall/F1 Scores of 84\% by using kNN classification.
\par
These fingerprints could be used to covertly monitor and track individuals carrying these devices. In a hypothetical situation, a person could be walking in a public space where Bluetooth devices are very common, and as a part of normal operation, their device emits Bluetooth signals. Equipped with a database of RF fingerprints that were collected in a controlled environment, we could deploy one or more SDRs and passively capture the Bluetooth signals from a distance, and afterward by continuously comparing the received signals with our RF fingerprint database, we could determine the location and movement patterns of the individual without their explicit consent as Givehchian et. al \cite{givehchian2022evaluating} has confirmed by tracking individual BLE devices in field tests (although with different fingerprint - CFO and I/Q offset).
\par
While our experimental study provides valuable insights about how to extract RF fingerprints and their effectiveness, it is important to acknowledge the need for further testing in real-world scenarios. Factors such as multipath interference and device movement can introduce signal amplitude variations that may affect the accuracy of our "scaling factor" fingerprint matching approach, on the other hand because of wearable devices internal temperature, CFO can change thus limiting our ability to classify the wearable device. Additionally, the presence of numerous transmitting devices in the environment can pose challenges to accurate classification At the current stage of our research, the distinction between device-specific and model-specific fingerprints remains unclear, with device-specific fingerprints carrying more significant privacy implications. It is also critical to expand the data set significantly to allow for more significant and broad results - this work is ongoing and we plan to publish a database of the most popular Bluetooth wearable devices at the end of the year 2023.

\section{Acknowledgments}
This research is funded by the Latvian Council of Science, project ``Automated wireless security analysis of wearable devices'' (WearSec), project No. lzp-2020/1-0395. We also thank the anonymous reviewers for their suggestions on how to improve this paper.

\balance
\bibliographystyle{abbrv}
\bibliography{sigproc}  
\end{document}